# New Pyrochlore Oxide Superconductor RbOs$_2$O$_6$

Shigeki Yonezawa*, Yuji Muraoka, Yoshitaka Matsushita and Zenji Hiroi

*Institute for Solid State Physics, University of Tokyo, Kashiwa, Chiba 277-8581*

**Abstract**

We report the discovery of a new pyrochlore oxide superconductor RbOs$_2$O$_6$. The compound crystallizes in the same β-pyrochlore structure as the recently discovered superconductor KOs$_2$O$_6$, where Os atoms form a corner-sharing tetrahedral network called the pyrochlore lattice with Rb or K atoms in the cage. Resistivity, magnetic susceptibility and specific heat measurements on polycrystalline samples evidence a bulk superconductivity with $T_c$ = 6.3 K.

KEYWORDS: superconductivity, pyrochlore oxide

Pyrochlore oxides have the general chemical formula A$_2$B$_2$O$_7$ or A$_2$B$_2$O$_6$O', where A is a larger cation and B is a smaller transition metal (TM) cation.[1] The ideal pyrochlore structure is composed of two types of cation-oxygen sublattices: one is a corner-sharing tetrahedral network composed of A atoms with an O' atom in the center of each tetrahedron, and the other is another tetrahedral network of B atoms with each B atom coordinated quasi-octahedrally by six O atoms. This type of tetrahedral network is called the pyrochlore lattice, and has been studied extensively in order to elucidate the effect of geometrical frustration on the properties of localized spin and itinerant electron systems.

Recently, superconductivity was found in Cd$_2$Re$_2$O$_7$ (Re$^{5+}$: 5$d^2$) at $T_c$ = 1 K for the first time in the family of pyrochlore oxides.[2-4] The mechanism of the superconductivity appears to be conventional, and may be understood in the framework of the weak-coupling Bardeen-Cooper-Schrieffer (BCS) theory.[5] Very recently, we have discovered a new pyrochlore oxide KOs$_2$O$_6$ which exhibits superconductivity at 9.6 K.[6] Our preliminary structural analysis indicates that it crystallizes in a cubic structure with space group $Fd\bar{3}m$, as in the ideal pyrochlore oxides, but with K atoms located at the O' site of the ideal pyrochlore structure. It is known that the pyrochlore structure sometimes tolerates vacancies at the A and O' sites.[1)] Such pyrochlore oxides are generally called defect pyrochlores. By contrast, $KOs_2O_6$ should not be classified as defect pyrochlores, because of the apparent difference in the metal occupations. Thus, we call this $AB_2O_6$ type oxide a β-pyrochlore oxide to distinguish it from ordinary defect pyrochlore oxides. In the following search for a new superconductor, we have obtained a new ternary phase $RbOs_2O_6$ with the β-pyrochlore structure, which exhibits superconductivity at 6.3 K.

Polycrystalline samples were prepared from $Rb_2O$ and Os. The two powders were mixed in the molar ratio of $Rb_2O$:Os = 1:4, ground and pressed into a pellet in a dry atmosphere. The pellet was heated in an evacuated silica tube at 773 K for 24 h. It was necessary in the preparation process to avoid the formation of highly toxic $OsO_4$. In order to control the oxygen partial pressure in the silica tube, an appropriate amount of AgO was added to the end of the silica tube: AgO decomposes into silver and oxygen above 370 K, and thus generates an oxidizing atmosphere. The chemical composition of the product examined by energy dispersive X-ray analysis in a scanning electron microscope was Rb:Os ~ 1:2.

Figure 1 shows a powder X-ray diffraction (XRD) pattern taken at room temperature. All the intense peaks can be indexed assuming a cubic unit cell with a lattice constant $a$ = 1.0114 nm. A few extra peaks from Os are also detected. Extinctions observed in the XRD pattern are consistent with the space group of $Fd\bar{3}m$, which is expected for the ideal pyrochlore structure. However, the relative peak intensities are significantly different from those of typical pyrochlore oxides, and are similar to those reported for $KOs_2O_6$.[6)] Therefore, it is thought that $RbOs_2O_6$ has the same β-pyrochlore structure as $KOs_2O_6$.

Resistivity measurements were carried out down to 0.5 K by the standard four-probe method in a Quantum Design PPMS equipped with a $^3$He refrigerator. As shown in Fig. 2(a), the temperature dependence of resistivity for $RbOs_2O_6$ exhibits good metallic behavior below room temperature, without any signs of phase transitions such as observed in $Cd_2Os_2O_7$[7)] or $Cd_2Re_2O_7$.[2)] It is also significantly different from that reported for $KOs_2O_6$: A clear $T^2$ temperature dependence is seen below 30 K for $RbOs_2O_6$, which is absent for $KOs_2O_6$.[6)] When a sample is further cooled, the resistivity sharply drops below 6.5 K due to superconductivity. The resistivity below the transition is nearly zero within our experimental resolution. The critical temperature $T_c$, defined as the midpoint temperature of the transition, is 6.3 K, and zero resistivity is attained below 6.1 K. It is to be noted that the resistivity starts to decrease significantly at the high temperature of about 8 K, although the reason for this is not clear.

In addition to the observation of the zero-resistivity transition, a large diamagnetic signal associated with the Meissner effect has been observed below 6.3 K. Figure 2(b) shows a temperature dependence of magnetic susceptibility measured on a powdered sample in a Quantum Design MPMS. The measurements were carried out in a magnetic field of 10 Oe on heating after zero-field cooling and then on cooling in the field. A superconducting volume fraction estimated at 2 K from the zero-field cooling experiment is almost 100 %, which is sufficiently large to constitute bulk superconductivity.

The superconducting transition has also been detected by specific heat $C$ measurements. As shown in Fig. 2(c), the specific heat divided by temperature suddenly increases below 6.3 K, and forms a broad maximum around 5.8 K. The shape of this anomaly is unusual, and is different from what one expects for a conventional superconductor. The details will be described elsewhere.

The superconductivity of $RbOs_2O_6$ is robust against magnetic fields as shown in the resistivity measurements under magnetic fields of Fig. 3(a). When the magnetic field is increased, the transition curve systematically shifts to lower temperatures. The superconductivity remains even at $\mu_0 H = 14$ T at 0.5 K. The field dependence of $T_c$, which was determined as the midpoint of the transition, is plotted in Fig. 3(b). The upper critical field at $T = 0$ may be around 17 T, which seems to be larger than Pauli's limit, 12 T, for a weak-coupling BCS type superconductor in the absence of spin-orbit interactions. However, as suggested by the previous band-structure calculations on related compounds,[8] the spin-orbit interactions can be significantly large in the 5$d$ TM pyrochlore oxides, and thus the actual Pauli's limit can be larger than 12 T.

In conclusion, we found superconductivity with $T_c = 6.3$ K in a new β-pyrochlore oxide $RbOs_2O_6$. The nature of this superconductivity is to be clarified in a future study. However, we believe that an interesting aspect of physics is involved in the superconductivity of $RbOs_2O_6$, as in that of $KOs_2O_6$, on the basis of the itinerant electrons on the pyrochlore lattice.

We thank F. Sakai for her help in the EDX analysis. This research was supported by a Grant-in-Aid for Scientific Research on Priority Areas (A) provided by the Ministry of Education, Culture, Sports, Science and Technology, Japan.

**Figure captions**

Fig. 1. Powder X-ray diffraction pattern of $RbOs_2O_6$. Peak index is given by assuming a cubic unit cell with a lattice constant *a* = 1.0114 nm. Asterisks mark extra peaks from Os.

Fig. 2. Temperature dependences of resistivity (a), magnetic susceptibility (b), and specific heat divided by temperature (c). Insets in (a) and (c) show enlargements around the superconducting transition. The magnetic susceptibility of (b) was measured on a powdered sample of $RbOs_2O_6$ in an applied field of 10 Oe. ZFC and FC indicate zero-field cooling and field cooling curves, respectively.

Fig. 3. (a) Temperature dependence of resistivity as a function of magnetic fields. (b) *H-T* phase diagram showing the temperature dependence of upper critical fields determined from the resistivity data shown in (a).

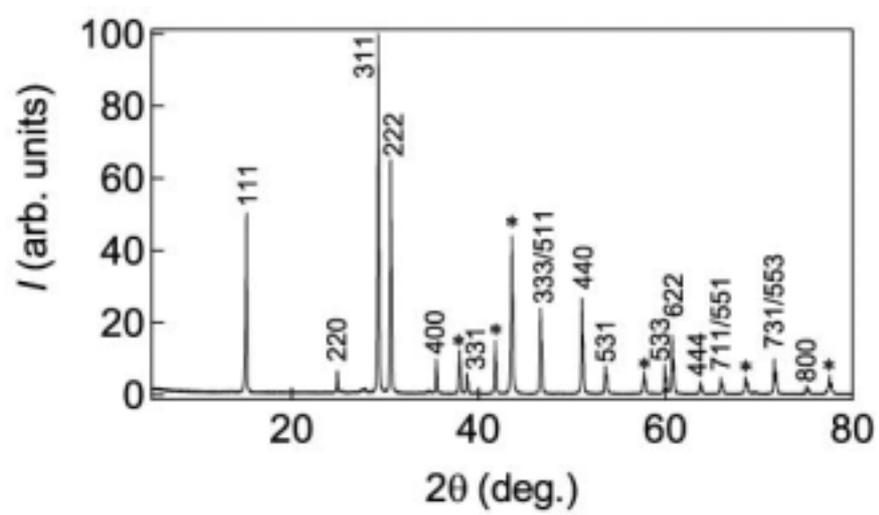

Fig.1

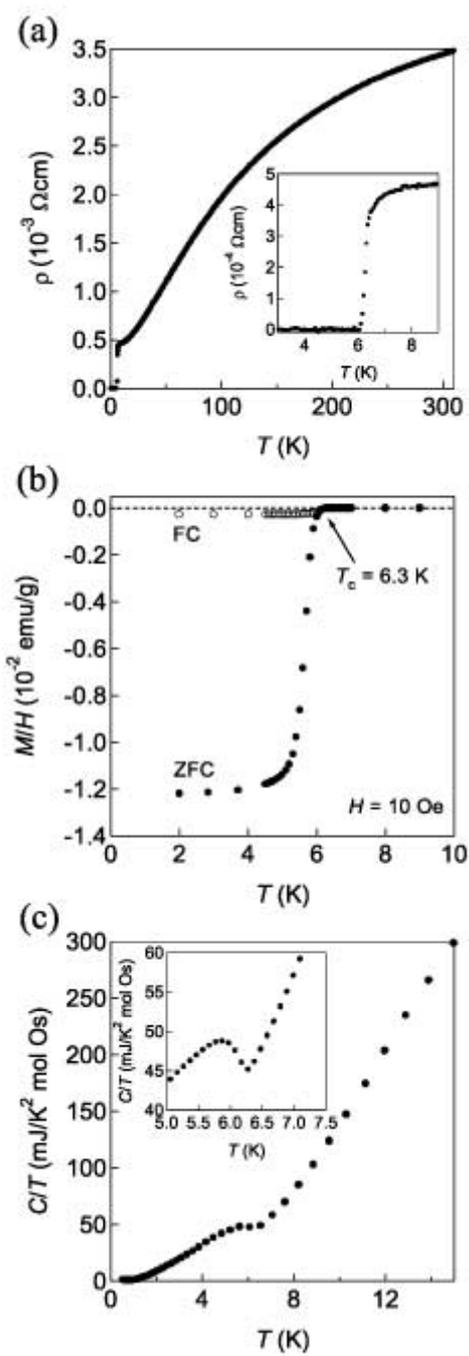

Fig. 2

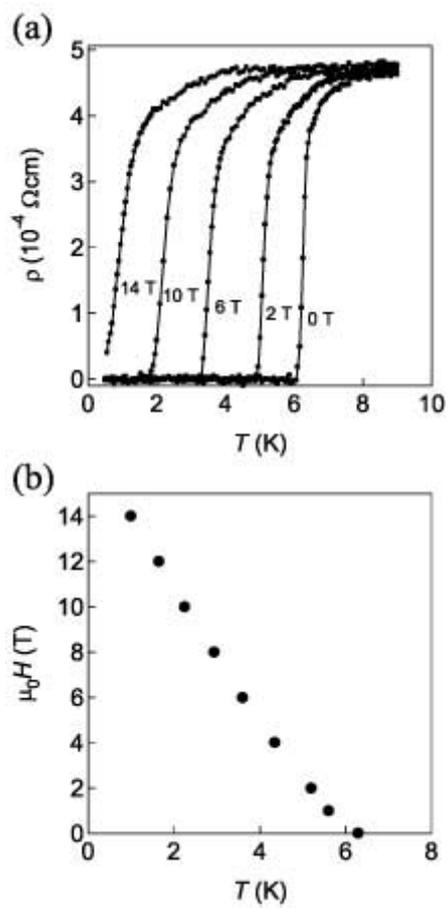

Fig. 3